# An efficient machine learning approach for extracting eSports players' distinguishing features and classifying their skill levels using symbolic transfer entropy and consensus nested cross validation

Amin Noroozi[1], Mohammad S. Hasan[1], Maryam Ravan[2], Elham Norouzi[3], and Ying-Ying Law[1]

Amin Noroozi * (corresponding author)
amin.noroozifakhabi@research.staffs.ac.uk
ORCID: 0000-0001-6937-6696

Mohammad S. Hasan
m.s.hasan@staffs.ac.uk

Maryam Ravan
mravan@nyit.edu
ORCID: 0000-0003-2055-9891

Elham Norouzi
elhamnorouzi85@gmail.com

Ying-Ying Law
ying-ying.law@staffs.ac.uk

[1] Department of Digital, Technologies and Arts, Staffordshire University, Staffordshire, England, UK
[2] Department of Electrical and Computer Engineering, New York Institute of Technology, New York, NY, USA
[3] Department of Computer Engineering, Azad University, Astara, Iran

**Acknowledgement**

"This preprint has not undergone peer review or any post-submission improvements or corrections. The Version of Record of this article is published in International Journal of Data Science and Analytics and is available online at https://link.springer.com/article/10.1007/s41060-024-00529-6 and can be viewed online at https://rdcu.be/dCIlA"

# Abstract

Discovering features that set elite players apart is of great significance for eSports coaches as it enables them to arrange a more effective training program focused on improving those features. Moreover, finding such features results in a better evaluation of eSports players' skills, which, besides coaches, is of interest for game developers to design games automatically adaptable to the players' expertise. Sensor data combined with machine learning have already proved effective in classifying eSports players. However, the existing methods do not provide sufficient information about features that distinguish high-skilled players. In this paper, we propose an efficient method to find these features and then use them to classify players' skill levels. We first apply a time window to extract the players' sensor data, including heart rate, hand activities, etc., before and after game events in the League of Legends game. We use the extracted segments and symbolic transfer entropy to calculate connectivity features between sensors. The most relevant features are then selected using the newly developed consensus nested cross validation method. These features, representing the harmony between body parts, are finally used to find the optimum window size and classify players' skills. The classification results demonstrate a significant improvement by achieving 90.1% accuracy. Also, connectivity features between players' gaze positions and keyboard, mouse, and hand activities were the most distinguishing features in classifying players' skills. The proposed method in this paper can be similarly applied to sportspeople's data and potentially revolutionize the training programs in both eSports and sports industries.

# 1 Introduction

eSports is a form of video gaming in which several players or teams compete against each other to achieve a predefined goal. The eSports industry has rapidly evolved from entertainment to a self-sustainable business within the last few years and still is actively developing in a variety of fields, including broadcasting, hardware, game statistics, streaming, and connectivity. The video game industry comprised over 80% of the total $36 billion profit earned from all software-related industries in 2018 in the United States [1]. Moreover, according to the global games market report, the gaming industry was worth over $180 billion in 2021 [1]. This fact gave rise to severe competition among game developers to provide a better user experience for the diverse population of video game players. This could be done using gameplay capable of being adaptively updated according to the players' skill levels, also referred to as dynamic difficulty adjustment (DDA) [2], which, in turn, entails an accurate evaluation of players' expertise first and foremost. Such an evaluation is of great value for eSports coaches as well. Currently, eSports coaches mostly rely on their knowledge and intuition for designing players' training programs. This is while in one of the most recent studies [3], professional eSports coaches counted the appraisal of players' skills and abilities as one of the biggest challenges during their careers. Having a reliable method to discover the features that distinguish high-skilled players could help coaches to design a more efficient program revolving around improving those features. Moreover, applying these features could improve the accuracy of players' skill classification, which is favourable for both game developers and eSports coaches.

Machine learning (ML) has been widely used for assessing eSports players' skills. In terms of the input data they use, the existing ML methods can be mainly categorized into two groups: methods using in-game data and methods using sensor data. The methods in the first group, accounting for a significant portion of the existing literature, mainly approach the problem from a statistical point of view and use in-game data and metrics such as Kill-Death Ratio (KDR), Win/Loss rate, etc., besides ML techniques to evaluate players' skill levels or predict the game's outcome [4–9]. These methods suffer from two major problems. Firstly, they have poor robustness due to being dependent on the games' features. When the game settings and features change, the previously trained model may not be valid anymore. Also, finding the correct in-game features capable of predicting the players' skill levels is not straightforward or even possible. Secondly, the main physical features that discriminate between players of different skill levels are not reflected in the game statistics. For example, it is not easy to comment on the players' reaction time, their skills in working with keyboards and mouse, and their cognitive abilities, such as sustained attention, only using game statistics. To address these problems, the methods in the second category have investigated the application of sensor data in assessing eSports athletes' abilities [10–13]. Different types of sensor data have already been used for this purpose in the literature, including gaze data [14], keyboard and mouse data [10], electroencephalography (EEG) [15], galvanic skin response (GSR) [11], skin resistance (SR) [11], heart rate [16], electromyography (EMG) [11], pupil diameter [11], face temperature [11], and players' movement data gathered by an accelerometer, magnetometer, and gyroscope [11, 12]. Methods in this category provide better robustness and accuracy as they are not dependent on the game's settings. However, they still do not provide adequate information about physical and cognitive features that set high-skilled players apart. The majority of these studies use sensor data gathered from only one part of the body [10, 14–16]. There are also a few studies using multiple sensor data [11–13]. However, the harmony between body parts is widely neglected in these studies. This is while such a factor could be decisive in differentiating between players of different skill levels in eSports and sports. For example, it is well established that the harmony between breathing and strides is one of the most distinguishing factors in professional runners [17].

The contributions of this paper can be summarized as follows: the first contribution of this paper is to propose an efficient ML algorithm to discover the features that distinguish professional players from amateur players. To our knowledge, this is the first study investigating this issue. We also use the selected features to increase the accuracy of players' skill classification. Secondly, the proposed method takes advantage of symbolic transfer entropy (STE) to extract connectivity features between sensors gathering data from different body parts such as eyes and hands. Although STE has been applied to sensor data of the same type, for example, EEG data, for other applications [18, 19], this is the first study applying STE to sensor data of different types for eSports players skill evaluation, thereby incorporating the harmony between body parts as features in the classification. Thirdly, we propose a novel feature selection procedure comprising a recently developed method called consensus nested cross-validation (CN-CV) [20] and the minimum redundancy maximum relevance (mRMR) method [21].

The rest of the paper is organized as follows: in section 2, we review the related research works presented in the literature. In section 3, we first introduce the dataset used in simulations and then describe the proposed methodology. The results and discussion are presented in sections 4 and 5, respectively.

## 2 Related work

Sensor data have been successfully applied to assessing eSports players' performance [22–27]. Gaze data gathered by eye-tracking sensors proved effective in differentiating between players of different skill levels. Investigators in [28] assessed the gaze behavior of amateur and professional players in the Counter-Strike: Global Offensive CS: GO game. Analyzing the gaze data of 15 players gathered by an EyeLink eye-tracker, they observed a considerable difference in the gaze position of players of different expertise. However, they did not report the classification results or any statistical measures for the observed differences. Similarly, a comparative study of amateur and professional players in CS: GO based on the gaze, keyboard, and mouse data is presented in [14]. For this purpose, the gaze behavior of 28 players, including 4 professional and 24 amateur players, was analyzed while the amateur players were divided into three groups: newbie, low-skill amateur, and high-skill amateur. Using the Mann-Whitney test, they observed a significant difference between the gaze behavior of professional and amateur players. For the classification, they divided the players into two classes, namely, professional and amateur. In the end, using a combination of keyboard, mouse, and gaze data, they reported a binary classification accuracy of 90%. However, due to the high degree of imbalance in their dataset, it is difficult to assess the validity of their results. One study [29] reported a significant difference in the players' distribution of visual fixation, gathered using a Tobii EyeX eye-tracker, in the CS: GO game. To achieve this result, they designed an experiment consisting of 21 players categorized into three groups: 10 newbies with less than 700 hours of playing experience, 7 amateur players with more than 700 hours of playing experience, and 4 professional players with more than 10000 hours of playing experience. Although they did not observe a significant difference in the mean visual fixation duration between players of different skill levels, the standard deviation of fixation duration was significantly different among players; players with higher skill levels had higher standard deviations. However, they did not assess the players' gaze location. Another study [30] presented an analysis of players' gaze data gathered by a Pupil Core eye-tracker in the StarCraft game. For this purpose, they used the gaze data of seven expert players and nine players with low skill levels while all participants played the game at three different difficulty levels. The gaze position of expert players covered a significantly larger interval in the horizontal direction compared to that of low-skilled players. Moreover, the intensity, number, and velocity of saccade in expert players were significantly greater than in low-skilled players.

Keyboard and mouse data have also been frequently used to assess the players' expertise level and performance in different games. A significant difference in keyboard and mouse usage between CS: GO amateur players and professional players was reported in [28]. They particularly considered the time interval in which the player used the A and D keys to move to the right and left and the time interval in which the player pressed the left mouse button (MOUSE1) while moving forward using the key W. In one study, the Mobalytics Proving Ground Assessment (MPGA) could differentiate between highly skilled and low-skilled League of Legends (LoL) players [31]. MPGA is an online assessment tool that evaluates LoL players' ability to work with a mouse and keyboard in response to randomly appearing targets. They first gathered keyboard & mouse data of 40 LoL players, including 20 highly skilled and 20 low-skilled players. After comparing the best scores of each group, they then found 17 variables significantly different between the two groups and concluded that MPGA could discriminate between high-skilled LoL players and low-skilled players successfully. Investigators in [32] recorded the mouse and keyboard data of 34 Red Eclipse game players divided into four classes based on their skill levels. Using features extracted from 60 seconds of the game, they reported a classification accuracy of 76%. However, they did not report the binary classification results. In another study [10], the mouse and keyboard data of 22 players, including 4 professional, 11 hardcore, and 7 casual amateur players, were gathered while playing the CS: GO game. A recursive feature selection technique was then used to determine the top 10 most important features for the characterization of the players. After analyzing the selected features, they found that the difference between amateur players and professional players could be characterized using the same features as the differences between amateur players of different skill levels. However, they also found that professional players of different skill levels could not be classified using the same features. Although they tried to alleviate the effect of imbalance in the dataset using a comparable number of amateur and professional players, they avoided classifying the players using the extracted features due to insufficient training data.

Several studies investigated the relationship between players' EEG data and their behavior and skills [15, 33, 34]. In [15], EEG data collected from 40 CS: GO players, including 20 amateur players and 20 professional players, were used to predict their skills and level of tiredness. Another study was conducted in [33] to discover the relationship between the players' EEG spectra and the selected moments of "win" and "lose" using the continuous wavelet transform. After decomposing the EEG signals of 10 players into five frequency bands of δ (1-4 Hz), θ (4-7 Hz), α (8-13 Hz), β (13-30 Hz), and γ (30-45 Hz), they observed a general decrease and increase in the alpha band for "loss" and "win" events, respectively. This is while they reported the opposite behavior in the theta band for the same events. A similar study is presented in [34] to discover the relationship between the players' EEG signals and their level of expertise in the Dota 1 game. For this purpose, they gathered EEG data of two players, including one newbie and one hardcore, while playing the game. In the end, they observed a considerable difference between the players' EEG signals in F7 and P3 electrodes according to the international 10-20 EEG placement.

Authors in [35] presented one of the first studies for analyzing the players' skills using sensor data. They initially recorded two players' GSR and SR data during different game events using a sensor glove. The gathered data together with players' reactions in different game situations were then fed to a Bayesian classifier to identify the player. The trained model could classify the players with 87% accuracy. However, despite the high accuracy, the sample size in their study was small, making it difficult to generalize the results to larger sample sizes. Moreover, their model still needs data related to the type of the players' reactions in the game, making it dependent on the gameplay. An evaluation of CS: GO players' skills using sensor data, including an accelerometer, magnetometer, and gyroscope is presented in [13]. Nineteen players, including 10 amateur and 9 professional athletes, were invited to play the game and their movement data were gathered using a set of sensors installed on a smart chair. The collected data for each player were then divided into 3-minute sessions resulting in 154 samples used for the classification. In the end, they reported an accuracy of 86% using the support vector machine (SVM) classifier. In another study [11], a similar experiment was conducted to evaluate LoL players' expertise using sensor data. For this purpose, they gathered sensor data of one amateur and one professional LoL team consisting of 10 players in total (5 amateur and 5 professional players), where each team played 11 matches over three experimental days. They split the gathered data into 3-minute sessions up to 5 sessions per match, resulting in 540 samples, which were finally used for the classification. The sensor features used for the prediction included GSR, heart rate, EMG, pupil diameter, frequencies of mouse clicks and keyboard strokes, gaze data, face temperature, oxygen saturation, and 3 features describing the players' movements. Finally, they could differentiate amateur players from professional players with 86% accuracy using the SVM classifier and the extracted features.

## 3 Methodology

### 3.1 Dataset

In this paper, we use a dataset presented by [12] and publicly available for download on GitHub [36]. The dataset is collected from 10 LoL players, organized into one amateur team and one professional team, including 5 amateur players and 5 professional players, respectively. Each team played up to 4 matches on 3 different days either versus bots or real opponents from the internet, while data from 12 sensors as well as environmental data such as temperature, pressure, altitude, humidity, and $CO_2$ level were gathered using a smart chair and a set of wearable sensors. The full description of the sensing system can be found in [12]. In total, 21 matches were played, 10 matches by professional players and 11 matches by amateur players. Therefore, up to $21 \times 5 \times 12 = 1260$ sensor data were gathered approximately, where each sensor data corresponds to the data gathered from a specific sensor attached to a specific player playing a specific match. After gathering the sensor data, the outliers for each sensor were removed, and then the signal was smoothed using an exponential moving average. Finally, all signals were resampled to a unified timestep of 1 second by averaging or summation, depending on the nature of the source sensor data. In addition to the sensor data, 2 other types of data were also gathered from in-game logs and surveys from each team member in each match. The in-game data, collected using the Riot API, include information about the key game events, including kill, death, and assist events, referred to as "moments of interest" (MoI) in this paper. The list of the sensor data used in simulations and their description are shown in Table 1.

## 3.2 STE

STE provides a robust, convenient, and computationally efficient measure to evaluate the flow of information in dynamic and multidimensional systems. This method is capable of measuring the strength and direction of information flow between time series recorded from structurally identical and nonidentical coupled systems (here, sensors of different types) using a technique known as symbolization. The STE is a directed measure. A large transfer entropy from *x* to *y* indicates that the past values of *x* help predict the current values of *y*, whereas a small transfer entropy indicates that the current value of *y* is independent of the past values of *x*.

Consider two time series $X = (x_1, x_2, ..., x_N)$ and $Y = (y_1, y_2, ..., y_N)$ where $x_i$ and $y_i$ are the $i^{th}$ time samples that are measured from two sensors. STE estimates the transfer of information between *X* and *Y* by symbolizing their recorded amplitude values. For this purpose, first for each given *i*, the *m* amplitudes $X_i = \{x_i; x_{i+d}; ...; x_{i+(m-1)d}\}$ are arranged in an ascending order as $X_i^a = \{x_{i+(k_{i1}-1)d} < x_{i+(k_{i2}-1)d} < \cdots < x_{i+(k_{im}-1)d}\}$ where *d* is the time delay and *m* is the embedding dimension, which shows the length of the comparing segments. A symbol sequence $\hat{X}_i = \{k_{i1}; k_{i2}; ...; k_{im}\}$ is then generated from $X_i^a$ where $k_{ij}, j = 1,2, ..., m$ are the indices of the elements of $X_i^a$. STE is then calculated from the two symbol sequences, $\hat{X}_i$ and $\hat{Y}_i$ as follows [37]

$$T_{Y,X}^S = \sum p(\hat{X}_{i+t}, \hat{X}_i, \hat{Y}_i) \log_2 \left( \frac{p(\hat{X}_{i+t}|\hat{X}_i, \hat{Y}_i)}{p(\hat{X}_{i+t}|\hat{X}_i)} \right) \quad (1)$$

In this paper, we use STE to measure the directed flow of information from the $i^{th}$ sensor to the $j^{th}$ sensor where $i, j \in \{1,2, ...,12\}$.

Table 1 List of sensor data used in simulations.

| Sensor data | Symbol | Description |
|---|---|---|
| Left hand movements | LHM | Data about hand movements were obtained by two inertial measurement units (IMU) placed on the left and right hands |
| Right hand movements | RHM | |
| Chair movements | CM | Chair movements were captured by an IMU sensor attached to the bottom of the chair |
| Gaze position | GP | The gaze position captured by the eye tracker |
| Pupil diameter | PD | The pupil diameter captured by the eye tracker |
| Electrodermal activity | EA | Data about the electrodermal activity are presented in terms of resistance measured in Ohms. |
| Left hand muscle activity | LHMA | Left hand muscle activity is measured by an EMG sensor and represented as voltage. |
| Right hand muscle activity | RHMA | Right hand muscle activity is measured by an EMG sensor and represented as voltage. |
| Heart rate | HR | Pulse data were collected by the heart rate monitor on the arm and measured in beats per minute |
| Keyboard activity | KA | Data about keyboard activity are presented as the number of buttons pressed in the last 5 seconds. |
| Mouse activity 1 | MA1 | The distance passed by mouse in the last 5 seconds |
| Mouse activity 2 | MA2 | The number of mouse clicks in the last 5 seconds |

## 3.3 Proposed ML algorithm

### 3.3.1 Data preprocessing

Using the entire recorded signal will include a significant amount of noise in calculations since not all parts of signals carry considerable information. The possible features that differentiate between players could be captured best when the players react to an event. For example, players' skills in working with keyboards could be observed better when they react to a game event. Otherwise, there is no considerable difference between players' keyboard data. In the LoL game, the key moments when players need to make decisions and react are Kill, Death, and Assist events, i.e., MoI. For this reason, we extract sensor data before, after, and at MoI using time windows of length $2t_d$, where the centers of the time windows for all sensor data are located at MoI. The rest of the data is then discarded, and only the extracted segments are used to calculate the STE features between different sensors. $t_d$ is a hyperparameter representing the delay in gathering the sensor data before and after each MoI.

Assume $t_e, e = 1, 2, \ldots, N$ and $T$ are MoI and the sampling rate, respectively. For a specific player playing a specific match, the extracted data for sensors $i$ and $j$ at each $t_e$, using the proposed time window of length $2t_d$, can be presented using two sequences containing $2S + 1$ samples as

$$X_i^e = (x_{t_e-ST}, \ldots, x_{t_e-2T}, x_{t_e-T}, x_{t_e}, x_{t_e+T}, x_{t_e+2T}, \ldots, x_{t_e+ST}) \tag{2a}$$

$$Y_j^e = (Y_{t_e-ST}, \ldots, Y_{t_e-2T}, Y_{t_e-T}, Y_{t_e}, Y_{t_e+T}, Y_{t_e+2T}, \ldots, Y_{t_e+ST}) \tag{2b}$$

where $t_d = ST$. In this study, the sampling rate is one second, $i.\,e.$, $T = 1$. Given $N$ MoI, data sequences containing all extracted samples at all MoI for sensors $i$ and $j$ can then be written as

$$X_i = (X_i^1, X_i^2, \ldots, X_i^e, \ldots, X_i^N) \tag{3a}$$

$$Y_j = (Y_j^1, Y_j^2, \ldots, Y_j^e, \ldots, Y_j^N) \tag{3b}$$

Sequences $X_i$ and $Y_j$ are then used to calculate the STE features between sensors $i$ and $j$.

MoI for each LoL game can be obtained using the RIOT API and are presented in the dataset. Figure 1 and Figure 2 show MoI on the heart rate data of one of the professional players and the proposed time window used for extracting the sensor data segments, respectively.

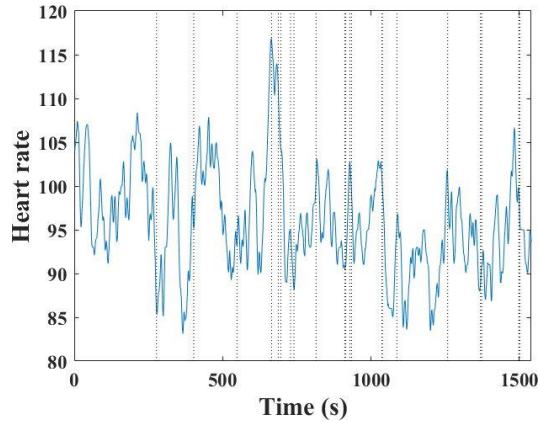

Fig. 1 MoI (dotted lines) and heart rate data (blue solid line) of one of professional players

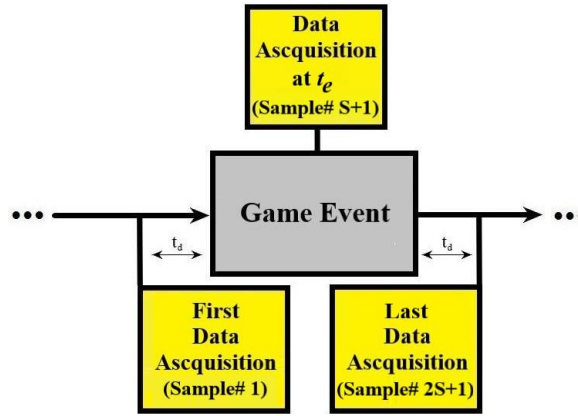

Fig. 2 The proposed time window to gather data at, before, and after MoI

### 3.3.2 Feature extraction

The STE feature $S_{ij}$ representing the connectivity between sensors $i$ and $j$ can be calculated using (1) and sequences of extracted data in (3a) and (3b). Since we use 12 sensors in simulations, the total number of STE features calculated for each player in each match is $N_c = 12 \times 12 = 144$. Whenever the number of events in a match was large, we broke the sequence of events into $K_s$ subsequences containing extracted data of between 4-10 events and then calculated the STE features. This resulted in $K_s$ samples of length $N_c$ for that match.

### 3.3.3 Feature selection

The number of extracted features from the previous stage is large, and it does not meet our final goal of finding the most distinguishing features that differentiate professional players from amateur players. Moreover, using a large number of features for classification will cause the underlying ML algorithm to overfit easily. For this reason, in this step, we select the $N_r$ most relevant features out of all $N_c$ features that are finally used for classification.

The existing feature selection methods, such as the mRMR technique [21], mainly rank features according to their importance. However, the optimum number of features for classification remains a challenge for scientists. This issue is grave for our problem as we need to determine the exact optimum features. To overcome this problem, we applied a feature selection procedure using the mRMR and CN-CV methods as follows:

**Step 1**: The training data are first divided into $K_{train}$ folds.
**Step 2**: One of the folds is removed, and the remaining $K_{all} - 1$ folds are merged into a fold, referred to as the outer training fold $K_{o_i}$ ($i = 1,2,3,\ldots,K_{train}$).
**Step 3**: The outer training fold $K_{o_i}$ is divided into $L$ inner fold.
**Step 4**: For each outer training fold $K_{o_i}$, the mRMR feature selection is applied to each of its inner folds $l$ ($l = 1, 2, \ldots, L$) to find the $N_i$ ($i = 1,2,3,\ldots,K_{train}$) most relevant features.
**Step 5**: Features with the highest frequencies across all inner folds are then selected for the related outer training fold $K_{o_i}$
**Step 6**: Repeat Steps 1-5 until all $K_{train}$ are removed once.
**Step 7**: The $N_r$ most common features (i. e. features with the highest frequencies) across all the $K_{o_i}$ outer folds are finally selected as the consensus features.

The CN-CV method proved computationally more efficient than its predecessor the nested cross-validation (NCV) since it selects the features without training a classifier. It also selects fewer irrelevant features compared to the NCV method [20]. Another advantage of CN-CV is that we do not need to be precise in setting $N_i$ in Step 4. If some irrelevant features are selected when setting $N_i$, they will be automatically removed when selecting the consensus features.

### 3.3.4 Classification and evaluation

In this step, the ML model was trained on the training data using the $N_r$ selected features, and enough care was taken to test the model on a separate unseen dataset. For this purpose, we used a $K_{all}$-fold cross validation (CV) to evaluate the model's classification performance on the entire dataset (including both testing data and training data). In this condition, one of the folds is reserved for testing, and the feature selection procedure will be applied to the remaining $K_{all} - 1$ folds, and this process continues until all folds are used for testing once. We evaluated the classification performance using three classifiers, including SVM, random forest (RF), and K-nearest neighbors (KNN). Applying these classifiers makes it easier to compare our results with previous studies that used the same dataset and classifiers [11].

### 3.3.5 The most distinguishing features

To find the most distinguishing features in the classification of eSports players' skill levels, we used two methods: in the first method, we applied the proposed feature selection to the entire dataset. In the second method, we used the results of the $K_{all}$-fold CV from the previous section. Applying the $K_{all}$-fold CV results in $N_{r_1}, N_{r_1}, N_{r_1}, \ldots, N_{r_{K_{all}}}$ selected features when the 1st, 2nd, 3rd, …, and $K_{all}$th fold is used for testing, respectively. The most distinguishing features were then found using the intersection of all $K_{all}$ feature sets as follows:

$$S = \cap_{j=1}^{K_{all}} S_j \tag{4}$$

where S is a set containing features existing in all $K_{all}$ CV iterations. $S_j$ is the feature set containing $N_{r_j}$ features selected in the $j$th iteration of the $K_{all}$-fold CV ($j = 1,2,3, ..., K_{all}$).

Computed features are not ranked as the CN-CV method is based on the frequency of features, not their rank. To get an ordered feature set, we additionally used the mRMR method to rank the features resulting from either of the methods mentioned above. The first method can be used when the only goal is to select the most discriminating features. The resulting features from this method cannot be used to test the classification performance, as no data are left for testing. On the other hand, the second method can be used when we need to evaluate the model performance and train it for prediction on an unseen dataset.

### 3.3.6 Setting $t_d$

To set hyperparameters $t_d$, we implemented an 8-step procedure as follows:

**Step 1**: Set $t_d = 1$ and calculate STE features. Split dataset into $K_{all}$ folds. These folds correspond to the same folds that will be used for the model evaluation in section 3.3.4.

**Step 2**: Remove one of the folds and merge the rest. Then split the merged data into 80% data for training and 20% data for validation.

**Step 3:** Apply the proposed feature selection method to 80% training data and use the selected features to train the model.

**Step 4:** Evaluate the trained model using the 20% validation data.

**Step 5:** Increase $t_d$ by a step value of one and repeat Steps 1-4 until $t_d = 10$.

**Step 6:** Select $t_d$ that produced the best results on the validation data.

**Step 7:** Repeat Steps 1-6 until all $K_{all}$ folds are removed once. This will produce $K_{all}$ values for $t_d$.

**Step 8:** Calculate the average $t_d$ over all $K_{all}$ folds.

The STE features are then recalculated using the selected $t_d$. These features were finally used for feature selection and classification.

The flowchart of the proposed algorithm, including all steps explained above, is presented in Figure 3.

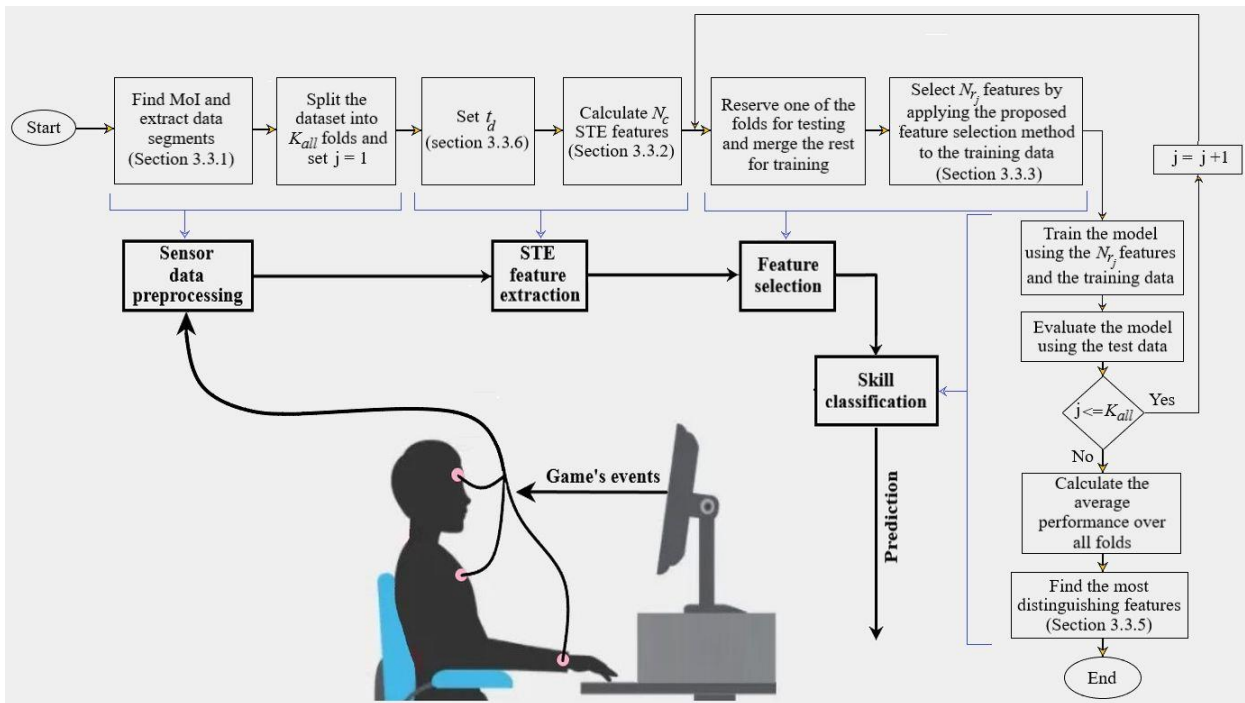

**Fig. 3** Flowchart of the proposed algorithm for finding the eSports players' most distinguishing features and classification of their skill levels

## 4 Results

In this section, we present the results of the proposed algorithm in finding the players most distinguishing features and classifying their skill levels. To prepare data, we extracted different events in each game played by each player and broke down the sequence of events into two subsequences ($K_s = 2$) for amateur players and three subsequences ($K_s = 3$) for professional players. This was because the number of events for professional players was higher than for amateur players. Therefore, each match for each amateur player and professional player resulted in 2 and 3 samples of length 144 respectively. Since 11 and 10 games were played by amateur players and professional players and there were 5 players in each match, this approximately resulted in up to 110 and 150 samples of length 144 for amateur players and professional players, respectively. However, to prevent the imbalance in the training dataset, we used 110 samples for both the amateur and professional classes.

### 4.1 The most distinguishing features

Table 2 shows the top 8 STE connectivity features selected from all 144 STE features. As can be seen, most of the selected features are related to the connectivity between the eye tracker sensor and other sensors, including mouse, keyboard, and hand activities. Therefore, our results suggest that the consistency between eyes and hand activities may be a decisive factor in distinguishing elite players from lower-rank players. The connectivity between keyboard and mouse activities, and between left hand and right hand activities are also among the selected features. This could indicate that professional players have better harmony between their left and right hands that work with keyboard and mouse, respectively.

Hand activity features, i.e. LHMA and RHMA, should not be mistaken with hand movement features, i.e. LHM and RHM. The LHMA and RHMA features represent muscle tension in players' reactions to the game events, but the LHM and RHM features represent players' hand movements while reacting to the game events. While connectivity between gaze position and hand muscle activities are among the top features, there are no features formed of hand movements. This shows that the difference in reaction time, which is the time lapse between observing an event and reacting to it, between professional and amateur players is more significant than the difference between how they react to the game events. This is also reflected in the mouse activity feature. While the number of mouse clicks is an element in features 2, 6, and 7, the distance passed by the mouse does not exist in any of the top features. This further emphasizes that the intensity and speed of players' reactions are more discriminating than how they react. Most players, whether they are professional or amateur, may know the correct movements and reactions. But whether they can swiftly apply their knowledge in a short time is a more decisive factor according to our results.

Figure 4 shows all samples in the dataset projected onto their first two principal components, PC1 and PC2. We removed a few outliers and normalized the principal components for a better visualization. As can be seen, the two classes are well separated using the principal components. Also, the number of professional players' data scattered among amateur players' data is more than the contrary. This shows that it is more likely to see a professional player underperforming in a match than an amateur player demonstrating a professional-level performance.

Table 2 Top eight the most distinguishing features in classification of players' skill levels

| Feature number | Selected connectivity (STE) feature between sensors |
|---|---|
| 1 | GP to KA |
| 2 | GP to MA2 |
| 3 | GP to LHMA |
| 4 | GP to RHMA |
| 5 | LHMA to RHMA |
| 6 | KA to MA2 |
| 7 | MA2 to KA |
| 8 | RHMA to LHMA |

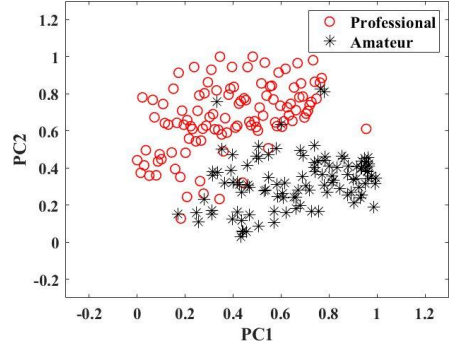

Fig. 4 The dataset projected onto the first two principal components, PC1 and PC2

## 4.2 Classification results

Figure 5 illustrates the average accuracy of three classifiers, namely, SVM, RF, and K- KNN, using a five-fold CV ($K_{all} = 5$) and different $t_d$ values. To compare the effect of different game events' data on the results, we additionally evaluated the classification performance when sensor data of only one game event, including kill, death, and assist, were used for classification individually. As can be seen, the best results were achieved using $t_d$ values of 4, 3, 4, and 4 when we used kill data, death data, assist data, and all events' data, respectively. In all cases, the proposed method with the optimum $t_d$ outperformed the previous method that reported a maximum accuracy of 85.6% using the same dataset [11]. As can be seen, for all classifiers, when the sensor data of all events are used for classification, the proposed method performs better than when only one event's data are used. This could be mainly due to the decrease in the number of training samples. Insufficient samples also affected the noise sensitivity of the algorithm such that the classifiers' performance started dropping at a smaller $t_d$ value (i.e., $t_d = 3$) when only death data were used for training (Figure 5b). The death event had the lowest frequency, and consequently, the lowest number of samples in the training dataset among other events. A similar effect can also be seen when other events' data are individually used for training, as the classification performance drops more abruptly in these cases by increasing $t_d$. Since we include noisier data with lower SNR in the training dataset by increasing $t_d$, this early drop in the performance could indicate that the results are impacted by noise more easily than when sufficient samples were used. The overall results using all events' data show that, on average, 5 seconds before and after each event, the effect of noise and irrelevant data prevails, and the sensor data at these points include negligible information about the players' skills. Also, the performance of classifiers does not rise and drop symmetrically with $t_d$ values smaller and greater than 4. The decrease in the classifiers' performance with $t_d > 4$ happens faster compared to its rise when $t_d < 4$. This in turn shows that data gathered at time points closer to the game events contain more relevant information about the players' skill than the data gathered at time points far from the game events. The accuracy of the classification using only kill and assist data is close, and the performance graphs show nearly the same pattern (Figure 5a and Figure 5c). This could be because the numbers of kill and assist events were close, and the EEG data fluctuations at these events were similar. Also, the overall performance looks more like the kill and assist classification performance graph than the death classification performance graph. This could mean that kill and assist events carry more information about the players' skills than death events.

Table 3 shows the classification performance, including accuracy, sensitivity, and specificity, for different classifiers and events' data using the optimum $t_d$ and a five-fold CV. The SVM classifier demonstrates the best performance in all cases. When all events' data were used for classification, SVM demonstrated an average accuracy of 90.3%, followed by RF and KNN, with average accuracies of 89.1% and 88.8%, respectively. The average performance of all classifiers is relatively close, which could be due to successful data preprocessing and feature extraction in the proposed method.

To further assess the robustness of the proposed method in dealing with unseen data, we evaluated the classification performance using the leave-one-subject-out cross-validation (LOSO-CV) method. For this purpose, at each iteration, we used one player's sensor data for testing while other players' sensor data were used for training the classifiers. Table 4 shows the results when all events' data were used for classification. Although the performance of all classifiers slightly dropped, their accuracies remain in an acceptable range, underscoring the good generalization ability of the proposed algorithm.

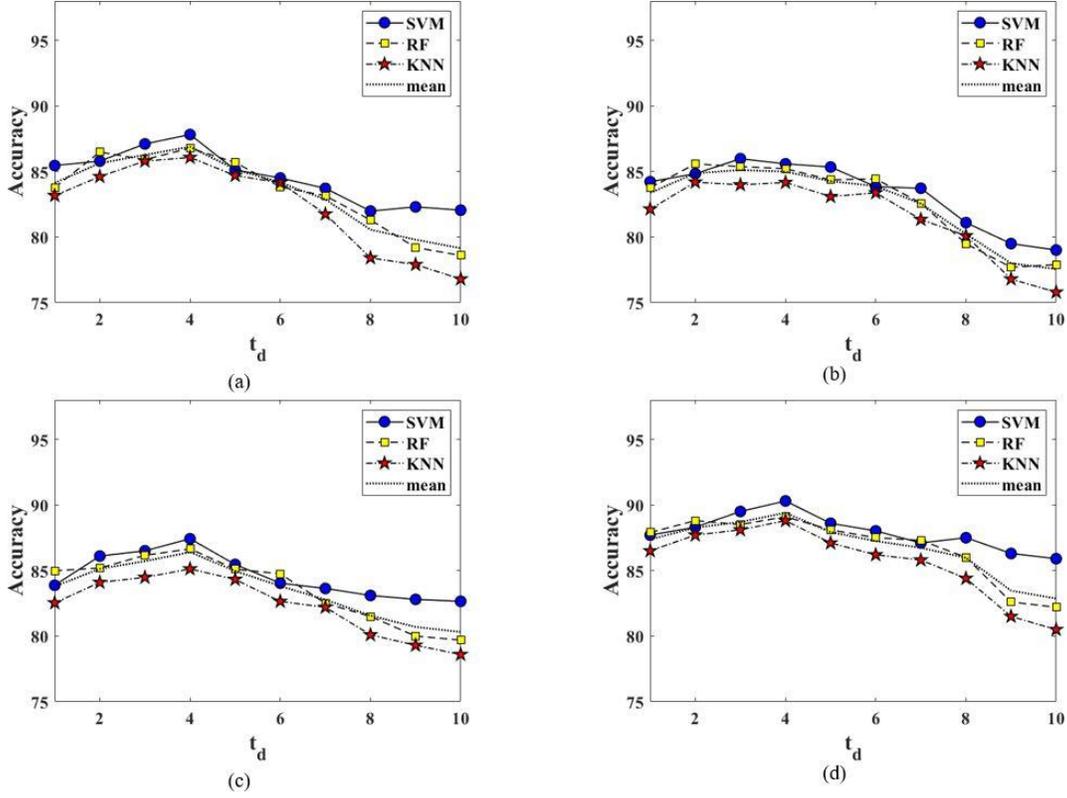

**Fig. 5** The average classification accuracy as well as the mean accuracy of all classifiers for different $t_d$ values when (a) only kill data, (b) only death data, (c) only assist data, and (d) all events' data were used. The accuracies are calculated using a five-fold CV

Table 3 Average accuracy of classification using a five-fold CV

| Classifier | Event data | $t_d$ | Accuracy | Sensitivity | Specificity |
|---|---|---|---|---|---|
| SVM | Kill | 4 | **87.8**%±2.6 | **85.3**%±2.6 | **90.6**±2.1 |
| RF |  |  | 86.7%±3.1 | 83.3%±2.2 | 90.1±2.4 |
| KNN |  |  | 86.1%±3.3 | 83.9±2.4 | 89.3±2.8 |
| SVM | Death | 3 | **85.9**%±2.2 | 83.1±2.9 | **88.2**±3.3 |
| RF |  |  | 85.3%±2.8 | **83.2**±2.7 | 87.4±2.6 |
| KNN |  |  | 83.9%±3.2 | 81.3±3.1 | 87.7±2.9 |
| SVM | Assist | 4 | **87.4**%±2.9 | **84.9**±1.9 | **91.6**±1.6 |
| RF |  |  | 86.6%±3.6 | 84.5±2.3 | 89.9±2.7 |
| KNN |  |  | 85.1%±2.9 | 82.1±2.2 | 88.7±2.5 |
| SVM | All events | 4 | **90.3**%±1.8 | **87.7**±2.1 | **95.6**±1.4 |
| RF |  |  | 89.1%±2.7 | 87.2±2.4 | 93.2±1.8 |
| KNN |  |  | 88.8%±2.4 | 87.4±3.1 | 91.1±2.1 |

Table 4 Average accuracy of classification using the LOSO-CV method

| Classifier | Accuracy | Sensitivity | Specificity |
|---|---|---|---|
| SVM | **87.5**%±2.2 | **85.6**±1.7 | **91.3**%±2.5 |
| RF | 85.1%±3.1 | 82.4±2.6 | 89.2%±1.9 |
| KNN | 82.3%±2.8 | 78.3±3.1 | 88.5%±2.4 |

## 5 Discussion

Sensor data proved to be effective in evaluating the skills of eSports players. However, there is not currently a reliable method to find the features that distinguish high-skilled players from low-skilled players. In this paper, we proposed a novel method to find these features and use them to classify the players into two groups: professional and amateur. For this purpose, we incorporated the relationship between different body parts into the classification stage using sensor data and the STE feature extraction method. To the best of our knowledge, this is the first study using connectivity features for the classification of eSports players.

The proposed method significantly outperformed the previous study using the same dataset and classifiers [11]. Also, the features representing the connectivity between the eye tracker sensor data, in particular the gaze position, and other sensor data, including keyboard, mouse, and hand activities, proved to be the most discriminating features between amateur players and professional players. These results suggest that the consistency between eyes and hands in professional players is a decisive factor distinguishing them from amateur players. Therefore, our results confirm the output of other studies [14, 28] that found a significant relationship between gaze position and the performance of players. However, these studies investigated the gaze data individually. According to our results, the connectivity between keyboard and mouse movements and between left and right hands are also among the top discriminating features in classifying players' skills. This, in turn, shows that the consistency and harmony between two hands working with the mouse and keyboard could play a significant role in the players' performance. Investigator [12] used the same dataset and found that mouse clicks and the distance passed by the mouse are among the features indicating players' higher chance of winning in the next encounter. However, according to our results, the distance passed by the mouse is not among the most important features. They also found that heart rate is a decisive factor, but our simulations do not show that heart rate has a significant effect on the classification performance. In the end, they achieved a low accuracy of 73.5% in predicting whether a player will lose the encounter occurring in 10 s. A similar research work is presented in [11]. In their study, a combination of 11 sensor data was employed to distinguish players of two skill levels, and a total classification accuracy of 85.6% was reported. However, in both previous studies, the gaze position was not among the most important features. In [11] they even observed a negligible correlation between the gaze positions of eSports players of the same skill level. This is mainly because the gaze position in their studies is considered individually without its relationship with other sensor data being factored in. Since the players' gaze positions may vary considerably according to the gameplay and the role they are playing, the correlations between different players' gaze data are expectedly minimal. This is while, according to our results, the gaze position emerges as an important factor when its connectivity with other sensor data is factored in.

The ML algorithm proposed in this paper is not dependent on the sensor type or game platform. Future work could apply the proposed method to other types of sensors in other games or applications, such as the evaluation of sportspeople's skills. For this purpose, the sensor data could be collected using different experiments, including evoked potentials test, alertness behaviour task (ABT), and concentration cognitive task (CCT). Examples of such datasets have already been presented in a few studies [38]. The high accuracy of the proposed method and its ability to extract the distinguishing features of professional players make it a promising tool for such applications.

# 6 Conclusion

We presented an efficient machine learning algorithm to find the features that distinguish professional players from amateur players in the LoL game and used them to classify the players' skill levels. For this purpose, we calculated STE features to incorporate harmony between body parts and then used the CN-CV method to select the most distinguishing features. The proposed method improved the classification accuracy compared to the previous study using the same dataset and classifiers and achieved an accuracy of 90.3%. The selected features provided meaningful insight into the characteristics setting professional players apart. Features representing the swiftness of players' reactions, such as connectivity between eyes (gaze positions) and hand muscle activities and between eyes and mouse and keyboard activities proved more decisive according to the results. The proposed method benefits game designers by providing better accuracy for DDA applications and will significantly help coaches design a more efficient training program for players. Since the proposed method in this paper is not dependent on the game settings and sensor types, it could also be applied to other video games and be a promising tool to evaluate sportspeople's skills. Therefore, this study could be regarded as a stepping stone to a future with bespoke training programs and intelligent game designs customized to the needs of players of different skills and abilities.

## Statements and Declarations

### Funding

The authors did not receive support from any organization for the submitted work.

### Competing interests

The authors have no relevant financial or non-financial interests to disclose.

### Compliance with ethical standards

The authors did not conduct any experiment involving human participants or animals and used a secondary dataset published in [12].

### Data availability statement

The dataset analyzed during the current study is available on GitHub at the following URL: https://github.com/smerdov/eSports_Sensors_Dataset

### Author contributions

The authors' contributions can be summarised as follows:
Conceptualization: Amin Noroozi and Ying-Ying Law; Methodology: Amin Noroozi, Mohammad S. Hasan, and Maryam Ravan; Formal analysis and investigation: Amin Noroozi, Mohammad S. Hasan, and Elham Noroozi; Visualization: Amin Noroozi and Elham Noroozi; Writing - original draft preparation: Amin Noroozi; Writing - review and editing: Amin Noroozi, Mohammad S. Hasan, and Maryam Ravan; Resources: Amin Noroozi and Maryam Ravan; Supervision: Mohammad S. Hasan and Ying-Ying Law